# Neutral-current neutrino-nucleus scattering off $^{127}$I and $^{133}$Cs: Coherent and incoherent contributions with electroweak refinements for odd-A nuclei

Muhammad Farooq[a,b], Shakeel Mahmood[a], Muhammad Faisal Khan[a]

a - Department of Physics, Air University, Islamabad, Pakistan

b - University of Caen Normandy, Caen, France

rmfchohan@gmail.com

shakeel.mahmood@au.edu.pk

2600437@students.au.edu.pk

**ABSTRACT:**

Calculations of neutral-current neutrino–nucleus scattering cross sections are essential for interpreting low-to-intermediate energy data, yet terrestrial measurements remain limited. The COHERENT Collaboration's 2017 observation of coherent elastic neutrino–nucleus scattering (CEvNS) on CsI[Na] at the Spallation Neutron Source reported 134 ± 22 (stat) ± 30 (syst) events, corresponding to ~77% of the Standard Model central prediction (173 ± 48), consistent within uncertainties but leaving room for improved theoretical modeling. Motivated by this, we compute neutral-current cross sections for $^{127}$I and $^{133}$Cs within a unified and self-consistent analytical framework that includes coherent elastic, incoherent excitation, and spin-dependent axial contributions for these odd-A nuclei, with a consistent distinction between nuclear and nucleon form factors. The incoherent component is treated following standard structure-function-based approaches, ensuring a physically motivated decomposition of the total cross section. Electroweak effects are incorporated through momentum-transfer-dependent $\sin^2\theta_W$ (MS scheme) and flavor-dependent neutrino charge radius corrections, consistently implemented in the vector couplings. Cross sections are presented as functions of neutrino energy and for decay-at-rest spectra, within the low-to-intermediate energy regime where elastic and quasi-elastic processes dominate. Inclusion of incoherent and axial terms enhances the total cross section by factors of $\sim 2$ at $E_\nu \approx 10\, MeV$ ($\sigma \approx 2 \times 10^{-42} cm^2$ versus pure CEvNS $\sim 10^{-42} cm^2$) and leads to incoherent dominance at $E_\nu \approx 50\, MeV$ ($\sigma \approx 10^{-40} cm^2$). Expected interaction rates for DAR neutrinos reach $\sim 0.1\ events/kg/yr$ near a $40\ keV$ recoil threshold. The results are benchmarked against CEvNS expectations and show qualitative consistency with COHERENT observations, while providing a systematic assessment of subleading contributions relevant for CsI-based detectors and astrophysical applications.

*Keywords:* Neutrino-nucleus scattering, Coherent elastic, Incoherent, Odd-A nuclei, $^{127}$I, $^{133}$Cs, COHERENT, Weak mixing angle, Neutrino charge radius

## 1. INTRODUCTION

Neutrino–nucleus interactions constitute a powerful probe of both electroweak dynamics and nuclear structure. Owing to their weakly interacting nature, neutrinos traverse nuclear matter with minimal disturbance, enabling scattering processes to access intrinsic nuclear properties such as nucleon distributions, spin configurations, and nuclear form factors. In particular, coherent elastic neutrino–nucleus scattering (CEvNS), first proposed by Freedman [1], has gained considerable attention following its experimental observation by the COHERENT Collaboration [2–5], owing to its enhanced sensitivity to weak neutral-current interactions. CEvNS occurs when the momentum transfer $q$ is sufficiently small such that the associated wavelength exceeds the nuclear size ($|q|R \ll 1$), where $R$ denotes the nuclear radius. In this regime, the scattering amplitudes of individual nucleons add coherently, leading to a cross section approximately proportional to the square of the neutron number, $\sigma \propto N^2 \, |F(q)|^2$, where $F(q)$ is the nuclear form factor [6,7] and $N$ is the neutron number of the nucleus, Neutrons dominate the weak nuclear charge because the proton contribution is suppressed by the weak mixing angle $\sin^2\theta_W$, resulting in the hierarchy:

$$|\, NQ_{\text{weak},n} \,| \gg |\, ZQ_{\text{weak},p} \,| \tag{1}$$

where $Z$ is Proton Number and $Q_{\text{weak},p/n}$ is Proton/Neutron Weak Charge. Consequently, CEvNS is primarily sensitive to neutron distributions inside the nucleus [6].

As the momentum transfer increases, the validity of the coherent description gradually diminishes. The nucleus can no longer be treated as a rigid object, and internal degrees of freedom begin to influence the scattering process. In particular, incoherent contributions arising from nuclear excitations, as well as spin-dependent (axial) effects, modify the cross section [6–8]. These effects are especially relevant for odd-mass nuclei such as Cesium-133 and Iodine-127, where non-zero nuclear spin enhances axial contributions. The suppression of the coherent component at higher momentum transfer (e.g., $q \sim 50$ MeV for $^{133}$Cs) further highlights the increasing importance of incoherent channels in this regime.

The distinction between coherent (elastic) and incoherent (inelastic) scattering was originally emphasized by Freedman [1], based on whether the nucleus remains in its ground state or undergoes internal excitation. This classification has since been revisited and extended in various theoretical frameworks, including beyond-the-Standard-Model interpretations [9–17]. Experimental confirmation of coherent elastic neutrino–nucleus scattering (CEvNS) has been reported by the COHERENT Collaboration [2–5], while related studies of nuclear response and energy transfer mechanisms have been informed by earlier investigations of X-ray and electron scattering on nuclei [18,19].

Most existing theoretical approaches treat either the purely coherent regime or incoherent, spin-dependent, and electromagnetic effects separately [9–17]. However, realistic detector materials and experimental conditions require a unified treatment valid across low and intermediate momentum-transfer regimes. Moreover, electroweak parameters entering the interaction are subject to radiative corrections. In particular, the weak mixing angle exhibits scale dependence, while neutrino electromagnetic properties—such as the effective charge radius $\langle r_{\nu_\alpha}^2 \rangle$and magnetic moment $\mu_\nu$—introduce additional corrections to the cross section. Although these effects are predicted to be extremely small within the Standard Model, their inclusion enhances theoretical precision and provides sensitivity to possible new physics beyond it.

Motivated by these considerations, the present work develops a unified analytical framework for neutral-current neutrino–nucleus scattering that consistently incorporates coherent, incoherent, spin-dependent, and electromagnetic contributions within a single formalism. In contrast to previous studies, the present approach includes: (i) nuclear excitation effects beyond the coherent limit, (ii) axial contributions relevant for odd-mass nuclei, and (iii) electromagnetic corrections arising from the neutrino charge radius and the momentum-transfer dependence of the weak mixing angle. The analysis is applied specifically to $^{133}$Cs and $^{127}$I targets, which are widely used in neutrino detection experiments [20–24] such as COHERENT, KIMS, NAIAD, DAMA/LIBRA, and SABRE.

This unified treatment provides a continuous description of scattering across different momentum-transfer regimes and leads to a modest modification of predicted cross sections and event rates. Such refinements are particularly relevant in light of experimental observations. For instance, the COHERENT Collaboration reported a CEvNS signal of $134 \pm 22$(stat) $\pm 30$(syst) events, compared to a Standard Model prediction of $173 \pm 48$[2–5], which is consistent within uncertainties but highlights the need for improved theoretical modeling beyond the simplest coherent approximation. In addition, tensions and uncertainties in related neutrino–nucleus processes, such as charged-current interactions on $^{127}$I, further motivate refined treatments across different interaction channels.

The remainder of this paper is organized as follows. Section 2 presents the theoretical framework and cross-section formalism. Section 3 discusses numerical results and interaction rates. Finally, Section 4 summarizes the conclusions.

## 2. MATHEMATICAL FRAMEWORK

### A. Cross Sections

The CEvNS differential cross section can be obtained from the general neutrino–nucleus scattering formalism in the coherent limit, where the momentum transfer is sufficiently small for all nucleons to contribute coherently. Under standard approximations, the cross section reduces to a form proportional to the square of the weak charge and the nuclear form factor, as discussed in Refs. [1, 25–27]. In particular, this expression corresponds to the simplified coherent cross section derived in Ref. [7], rewritten in terms of the recoil kinetic energy $E_T$ and incident neutrino energy $E_{v_\alpha}$ ($\alpha = e - electron, \mu -\ muon,\ \tau - tau$) and the notation adopted in this work:

$$\frac{d\sigma_c}{dE_T} = \left[C_1\left(1 - \frac{E_T}{E_{v_\alpha}}\right) - C_2\frac{E_T}{E_{v_\alpha}^2}\right] Q_w^2 F_w^2(q^2) \tag{2}$$

where

$$c_1 = \frac{G_F^2 m_A}{\pi} \qquad\qquad c_2 = \frac{c_1 m_A}{2} \tag{3}$$

$G_F$ and $m_A$ are Fermi constant and the mass of nucleus, respectively. The Nuclear weak charge $Q_w$ can be given [28] in terms of the number of nucleons:

$$Q_w = g_n^V N + g_p^V Z \tag{4}$$

The weak vector couplings ($g_{p/n}^V$) of the proton and neutron is given by the Standard Model as

$$g_p^V = \frac{1}{2} - 2Sin^2\theta_w, \qquad g_n^V = -\frac{1}{2} \tag{5}$$

and the nucleus weak form factor ($F_w(q)$) can be given by;

$$F_w(q) = \frac{1}{Q_w}\left[ZQ_{\text{weak},p}F_p + NQ_{\text{weak},n}F_n\right] \tag{6}$$

where $F_{p/n}$ is the proton/neutron form factor. Generally, the proton contribution is ignored due to its small weak charge, but it is not negligible at high momentum transfer, particularly when incorporating spin effects, incoherent contributions, and electromagnetic interactions into the cross section.

The momentum transfer can be expressed in terms of the recoil kinetic energy as:

$$q = (2m_A E_T)^{\frac{1}{2}} \tag{7}$$

This relation is obtained under the assumptions of elastic scattering and non-relativistic nuclear recoil ($E_T \ll m_A$), where the nucleus remains approximately at rest before the interaction. These conditions are well satisfied for neutrino scattering considered in this work.

A best-fit value of $\sin^2\theta_W = 0.2312$ is used in cross-section estimations, which was established using the MS renormalization scheme and low-energy neutrino data [29]. We further discuss its dependence on charge radius and momentum transfer.

The spin contribution ($\sigma_s$) is a part of neutrino–nucleus scattering cross sections, acting as a special correction [7] in the general differential cross-section formula, which is important for consideration when dealing with even-odd proton–neutron nuclei. The modified form for the spin contribution for odd-A nuclei can be given by:

$$\frac{d\sigma_s}{dE_T} = \frac{1}{8}\left[1-(-1)^{N+Z}\right]\left[C_1\left(1-\frac{E_T}{E_{v_\alpha}}\right)+C_2\frac{E_T}{E_{v_\alpha}^2}\right] Q_A F_A(q) \tag{8}$$

Here, $Q_A$ denotes the effective axial charge of the nucleus, which depends on the nuclear spin structure and $F_A(q)$ is the nucleon axial form factor. This contribution vanishes for spin-less nuclei having even numbers of protons and neutrons but plays a vital role for odd-A nuclei having non-zero spin. The factor $[1 - (-1)^{N+Z}]$ acts as a selection rule: it vanishes for even–even nuclei ($N + Z$ even), corresponding to spin-zero systems, and becomes non-zero for odd-A nuclei where the nuclear spin is non-vanishing. At low momentum transfer, the spin-dependent contribution is much smaller than the coherent (vector) contribution. However, at higher momentum transfer, the coherent cross section is suppressed by the nuclear form factor $F_w(q^2)$, making the relative contribution of the spin-dependent term more significant.

Moreover, the discussion of incoherent cross-sections becomes important at high values of momentum transfer $(q)$ due to an increase in the probability of inelastic interactions, where the neutrino deposits energy into the nucleus. The interaction in this case may excite one or more nucleons or nucleon clusters within the nucleus. The nucleus may or may not conserve its integrity. The regime discussed here assumes conserved nuclear integrity. The recoil kinetic energy of the nucleus ultimately increases with $E_{\nu_\alpha}$. At finite momentum transfer, the neutrino probes individual nucleons inside the nucleus, and the scattering amplitude can be expressed as a sum over nucleon contributions. The corresponding cross section involves nuclear density correlations, which naturally separate into coherent and incoherent parts. The coherent contribution arises from the average nuclear density and scales with $|F(q)|^2$, while the incoherent contribution originates from fluctuations around this average and is proportional to $1 - |F(q)|^2$.

This decomposition is well established in the literature (see, e.g., Refs. [25–27]). In the present work, we follow this formalism and adopt the expressions for the incoherent cross section accordingly. The results are rewritten in terms of the notation used here, without introducing additional approximations. Under these considerations, the proton and neutron contributions to incoherent scattering can be written as:

$$\frac{d\sigma_{ip}}{dE_T} = 2c_1(1 - |F_p|^2) \times Z\left((g_p^V)^2 + (g_p^A)^2 + (g_p^R)^2 \frac{E_T}{E_{\nu_\alpha}}(\frac{E_T}{E_{\nu_\alpha}} - 2) - 2g_p^L g_p^R\left(\frac{E_T m_p}{2E_{\nu_\alpha}^2}\right)\right) \tag{9}$$

$$\frac{d\sigma_{in}}{dE_T} = 2c_1(1 - |F_n|^2) \times N\left((g_n^V)^2 + (g_n^A)^2 + (g_n^R)^2 \frac{E_T}{E_{\nu_\alpha}}(\frac{E_T}{E_{\nu_\alpha}} - 2) - 2g_n^L g_n^R\left(\frac{E_T m_n}{2E_{\nu_\alpha}^2}\right)\right) \tag{10}$$

where $g_{p/n}^{R/L}$ is right/left chirality proton/neutron coupling, $g_{p/n}^A$ is proton/neutron axial coupling and $m_{p/n}$ is proton/neutron mass. The net incoherent part of the neutrino–nucleus differential cross section, which is important to discuss for higher values of momentum transfer, can be given by the sum of both contributions.

The right- and left-chirality couplings can be expressed in terms of the vector and axial couplings, and they play a vital role in determining the incoherent contribution in neutrino–nucleus scattering:

$$g_{p/n}^L = \frac{1}{2}(g_{p/n}^V + g_{p/n}^A) \qquad g_{p/n}^R = \frac{1}{2}(g_{p/n}^V - g_{p/n}^A) \tag{11}$$

In this work, a simplified treatment of the axial couplings is adopted, where nucleon-level values are used:

$$g_p^A = \frac{1}{2} \qquad\qquad g_n^A = -\frac{1}{2} \tag{12}$$

A complete treatment would require detailed nuclear structure calculations involving spin-dependent nuclear response functions. However, since the present analysis is dominated by the coherent (vector) contribution, this approximation is sufficient for estimating the axial effects.

In addition to the SM weak interactions, neutrinos may interact via their magnetic moment. This is an electromagnetic interaction and requires a neutrino spin flip, which is why it adds incoherently to the SM cross section. The contribution [30] can be given by:

$$\frac{d\sigma_{EM}}{dE_T} = \frac{\pi\alpha_f^2\mu_\nu^2\, Z^2}{m_e^2\, E_{\nu_\alpha}}\left(\frac{E_{\nu_\alpha} - E_T}{E_T} - \frac{E_T}{4E_{\nu_\alpha}}\right)F_{ch}(q^2) \tag{13}$$

It has a $1/E_T$ dependence on the recoil energy $E_T$, making the scattering cross section more pronounced at low energies. $m_e$ is the electron mass, $F_{ch}(q^2)$ is the nuclear charge form factor, which can be parameterized in a similar manner to the weak form factor but with neutron contribution vanishing, and $\alpha_f$ is the fine-structure constant, on which we show dependence of the cross-section.

The EM term is completely independent of the weak interaction, so it adds incoherently. Its magnitude is directly proportional to $\mu_\nu^2$ , Although $\mu_\nu$ is very small, it can enhance the cross section, especially at low recoil energies, which is relevant for low-energy thresholds and high-resolution detectors. Experimentally, any enhancement above the SM prediction in the low-energy recoil spectrum signals physics beyond the Standard Model.

A general net cross section can be given by the sum of all contributions. The contributions can vanish for specific conditions. The solution implies better results when studied and it allows us to study the regime shift in high/low energy transfer. The dependence factors and details about the regimes are given ahead. The net differential cross section can be integrated for a detector having threshold energy ($E_T^{th}$) and the maximum detectable kinetic energy ($E_T^{\max}$):

$$\sigma(E_{\nu_\alpha}) = \int_{E_T^{th}}^{E_T^{\max}} \frac{d\sigma}{dE_T}\, dE_T \tag{14}$$

The maximum recoil kinetic energy of the nucleus can be given in terms of the incoming neutrino-energies:

$$E_T^{max}\left(E_{\nu,\alpha}\right) = \frac{2E_{\nu_\alpha}^2}{m_A + 2E_{\nu_\alpha}} \tag{15}$$

**B. Form Factors**

The derivation of the exact nuclear form factor is complex, so empirical fits such as the Helm model [31] can be used as the nuclear form factor.

$$F_{helm}(q) = \frac{3j_1(qR)}{qR}\exp\left(-\frac{q^2s^2}{2}\right) \tag{16}$$

With $s$ denotes the nuclear skin thickness parameter and $j_1(x)$ is spherical Bessel function of order one. The Helm form factor can also serve as the charge form factor in electromagnetic interactions.

The form factor approaches unity at zero momentum transfer and approaches zero at larger momentum transfer. In the coherent neutrino–nucleus scattering system, the nucleus acts as an isolated particle, for which the Helm form factor can be used by plugging in the constants associated with the nucleus.

The quantities $F_{\mathrm{p}}(q)$ and $F_n(q)$ represent nuclear form factors, corresponding to the spatial distributions of protons and neutrons inside the nucleus, and are modeled using the Helm parameterization [31], by incorporating the skin thickness parameter $s_{p/n}$ and Radius $R^{p/n}$ corresponding to the proton/neutron:

$$F_{p/n}(q) = \frac{3j_1\left(qR^{p/n}\right)}{qR^{p/n}}\exp\left(-\frac{q^2s_{p/n}^2}{2}\right) \tag{17}$$

Another analytical expression for the nuclear form factor is provided by the Symmetrized Fermi (SF) model [32,33]. While the Helm form factor also admits a simple analytic expression, the SF form factor is derived directly from a realistic symmetrized nuclear density distribution. As a result, it provides an exact analytic connection between the density parameters and all moments of the distribution without relying on additional approximations. This makes it particularly suitable when higher-order moments are relevant for the analysis. The form factor $F_{SF}(q)$ can be expanded as a Taylor series [32] at low momentum transfers:

$$F_{SF}(q) = 1 - \frac{q^2}{3!}R^2 + \frac{q^4}{5!}R^4 - \frac{q^6}{7!}R^6 + \cdots \tag{18}$$

The first three moments of the Symmetrized Fermi Form factor can be given as:

$$\begin{aligned}
R^2 &\equiv \langle r^2\rangle = \frac{3}{5}c^2 + \frac{7}{5}(\pi a)^2 \\
R^4 &\equiv \langle r^4\rangle = \frac{3}{7}c^4 + \frac{18}{7}(\pi a)^2c^2 + \frac{31}{7}(\pi a)^4 \\
R^6 &\equiv \langle r^6\rangle = \frac{1}{3}c^6 + \frac{11}{3}(\pi a)^2c^4 + \frac{239}{15}(\pi a)^4c^2 + \frac{127}{5}(\pi a)^6
\end{aligned} \tag{19}$$

where $c$ is the half-density radius and $a$ is the surface diffuseness parameter of the nuclear density distribution.

For greater q-value, we need to use exact parameterization for the SF form factor [32]:

$$F_{SF}(q) = \frac{3}{qc((qc)^2 + (\pi qa)^2)}\left(\frac{\pi qa}{\mathrm{Sinh}[\pi qa]}\right)\left(\frac{\pi qa}{\mathrm{Tanh}[\pi qa]}\mathrm{Sin}[qc] - qc\mathrm{Cos}[qc]\right) \tag{20}$$

In this formulation, the oscillatory behavior of the form factor is governed by the half-density radius $c$, while the exponential falloff at large momentum transfer is controlled by the surface diffuseness parameter $a$.

The COHERENT Collaboration has adopted the Klein–Nystrand (KN) form factor [34] to describe the nuclear charge distribution relevant to their experimental studies of neutrino interactions. This form factor is obtained through a convolution process involving a Yukawa potential, characterized by a range parameter $a_k = 0.7\ fm$, and a Woods–Saxon nuclear density distribution. The Yukawa potential models the spatial extent of the force, and the Woods–Saxon distribution provides a realistic representation of the nuclear density profile that smoothly decreases at the nuclear surface.

A hard-sphere model is used to simplify the mathematical treatment and enhance computational efficiency, approximating the Woods–Saxon distribution. In this approximation, the density of nuclear matter is assumed to be uniform throughout a sphere of radius $R$, with a sudden drop to zero outside this radius. Upon convolution of the Yukawa potential with this simplified density, the KN form factor obtained turns out to be very effective in encompassing most significant features of nuclear structure related to coherent neutrino–nucleus scattering. This form factor is critical for theoretical calculations and experimental interpretations in the COHERENT Collaboration's work and enables modeling of interaction processes with high accuracy,

$$F_{KN}(q) = \frac{3j_1(qR)}{qR}\left(\frac{1}{1 + (qa_k)^2}\right) \tag{21}$$

where $a_k$ is a parameter describing range of the interaction.

We note that the axial contribution in neutrino–nucleus scattering is, in general, governed by nuclear spin-dependent structure functions. A complete treatment would require detailed nuclear structure calculations. However, for the present analysis, which focuses primarily on the coherent (vector) contribution, the axial term is expected to be subdominant. Therefore, we approximate it using the nucleon axial form factor to provide an order-of-magnitude estimate. The Axial form factor for the nucleon current can be given by an interpolated function [35] with a modification constant for the neutrino-nucleus scattering:

$$F_A(x) = \frac{F_{A0}}{1 + x^2}$$
$$x = \frac{q^2}{m_{p/n}^2} \tag{22}$$

where $m_{p/n}$ is mass of the nucleon. The Neutral Current Weak Axial charge ($F_{A0}$) can be well-approximated:

$$F_{A0} \approx -0.687 \tag{23}$$

### C. Spectral Functions and Event Rates

Secondary pions are produced in large quantities when a high-intensity proton beam in the few-hundred MeV to approximately 1 GeV range strikes a dense target. The creation of a clean stopped-pion neutrino source is made possible by the strong suppression of the pion decay-in-flight component for proton energies below roughly 1 GeV. While $\pi^-$ are mostly absorbed by nuclei, most $\pi^+$ in a dense target lose energy, stop, and decay at rest.

The prompt two-body decay produces the neutrino yield from pions with a monoenergetic 29.8 MeV $\nu_\mu$, followed by the delayed three-body muon decay giving well-defined $\nu_e$ and $\bar{\nu}_\mu$ spectra up to $E_\nu^{max} = m_\mu/2 \approx 52.8$ MeV with $m_\mu$ the muon mass.

The corresponding Spectral Functions [36] are given by:

$$\begin{aligned}
\Phi_{\nu_\mu}(E_\nu) &= \frac{2m_\pi}{m_\pi^2 - m_\mu^2}\,\delta\left(1 - \frac{2E_\nu m_\pi}{m_\pi^2 - m_\mu^2}\right) \\
\Phi_{\nu_e}(E_\nu) &= \frac{192}{m_\mu}\left(\frac{E_\nu}{m_\mu}\right)^2\left(\frac{1}{2} - \frac{E_\nu}{m_\mu}\right) \\
\Phi_{\bar{\nu}_\mu}(E_\nu) &= \frac{64}{m_\mu}\left(\frac{E_\nu}{m_\mu}\right)^2\left(\frac{3}{4} - \frac{E_\nu}{m_\mu}\right)
\end{aligned} \tag{24}$$

These spectral functions can be used with the differential cross section to find the double differential event rate [37]:

$$K\,(E_T) = \frac{d^2N}{dE_T\,dE_{\nu_\alpha}} = \frac{N_A\ \times 10^3}{A_t} \times \frac{d\sigma}{dE_T}\,(E_\nu, E_T) \times \Phi_\nu\,(E_\nu) \tag{25}$$

Here $N_A$is Avogadro's number, the factor $10^3$ converts kg↔g if needed, and $A_t$ is the target mass number. Integrating over $E_\nu$ while keeping $E_T$ fixed gives single-differential recoil spectrum where Integrating over $E_T$ but keeping $E_\nu$ fixed gives events per neutrino energy. We use above spectral functions $\Phi(E_\nu)$ in our calculations in order to interpret the interesting results.

### D. Neutrino Charge Radius, $\boldsymbol{sin^2\,\theta_W}$ and Fine Structure Constant

One-loop electroweak corrections to the photon–neutrino vertex, namely the charged-lepton–W boson loop, are the source of the neutrino charge radius in the Standard Model. This quantity enters through an effective shift of the vector coupling. The charge radius, which is not a physical size but rather an electromagnetic radiative effect, is a gauge-invariant combination of vertex and external leg corrections. The expression for its Standard Model value [38] is

$$\langle r_{\nu_\alpha}^2 \rangle = \frac{G_F}{4\sqrt{2}\pi^2}\left[3 - 2\log\frac{m_\alpha^2}{m_W^2}\right] \tag{26}$$

Where $m_W$ is the mass of W-Boson and $m_\alpha$ is the charged lepton mass. The neutrino charge radius represents the neutrino's effective size in electroweak interactions. Its presence causes a shift in the effective weak mixing angle by altering the amplitude of neutral-current interactions. This shift can change the strength of the vector coupling in neutrino–nucleus neutral-current scattering by increasing or decreasing the effect, depending on the convention used. This matters for testing the Standard Model's consistency and for precision measurements. A more thorough explanation of neutrino electroweak properties can be obtained by incorporating the charge radius into calculations, which also facilitates comparison of theoretical predictions with experimental observations. In this work, the charge radius is evaluated separately for each neutrino flavor and incorporated into the cross-section calculations. Its impact on the CEvNS cross section is illustrated in Fig. 8 and subsequent figures, where flavor-dependent deviations are observed. Moreover, the weak mixing angle has some dependence on momentum transfer, which vanishes at high q values, as shown by the following parametrization:

$$\sin^2\theta_W \rightarrow (\sin^2\theta_W)_{eff} = 0.2312 + \frac{0.2381 - 0.2312}{1 + q^2/m_Z^2} + \frac{\sqrt{2}\pi\alpha_f}{3G_F}\langle r_\alpha^2\rangle \tag{27}$$

Where $m_Z$ is the mass of Z-Boson.

The running of the electromagnetic coupling arises from vacuum polarization effects, leading to a scale-dependent effective coupling. In the present work, rather than implementing the full dispersive treatment of vacuum polarization, we adopt a phenomenological parametrization designed to reproduce the known qualitative and quantitative behavior (as in [39-41]) over the momentum-transfer range relevant for neutrino scattering.

At all energy scales, the fine-structure constant, also known as the electromagnetic coupling constant, is not actually constant. The effective value of the coupling changes with momentum transfer because virtual charged particle–antiparticle pairs screen the electric charge due to vacuum polarization effects. The fine-structure constant has its well-known value of about 1/137 at low energies, but it increases to about 1/128 at higher energies as it approaches the scale of the Z boson mass. When calculating momentum-dependent processes, such as neutrino scattering where the exchanged boson probes various energy scales, this running behavior is a crucial component of precision electroweak physics. We show the dependence using the following parameterization:

$$\alpha_f = \frac{1}{137}\left(1 + 0.06\frac{q^2}{q^2 + (1\ GeV)^2}\right) \tag{28}$$

This form provides a smooth interpolation between the low-energy limit, where vacuum polarization effects are suppressed, and the high-energy regime, where screening leads to an effective increase in the electromagnetic coupling. While approximate, this parametrization captures the dominant scale dependence and is sufficient for the level of precision considered in this analysis.

The vector couplings in neutral-current interactions are directly impacted by modifications to the effective Weinberg angle. Any change in $\sin^2\theta_W$ affects the relative strength of neutrino interactions with protons and neutrons because the vector couplings explicitly depend on it.

Consequently, coherent and incoherent neutrino–nucleus scattering cross sections are altered. Even a slight variation in the vector coupling can significantly alter the total cross section in coherent scattering, where contributions from every nucleon add constructively. In incoherent processes, where contributions from individual nucleons, impact energy, and angular distributions, accurate predictions of neutrino–nucleus interaction rates must account for these effects.

The neutrino charge radius is flavor dependent through its dependence on the charged lepton mass in the loop. In the present work, this dependence is implemented explicitly by using the corresponding charged lepton mass for each neutrino flavor considered. In particular, for electron neutrinos and muon antineutrinos, the charge radius is evaluated using $m_e$ and $m_\mu$, respectively. This distinction propagates into the effective vector coupling and consequently affects the computed cross sections. The results, shown in Fig. 8 and above, therefore consistently incorporate flavor-dependent electroweak corrections.

The neutrino charge radius contribution is included as an additive shift to the vector coupling, while the running of the weak mixing angle accounts for universal electroweak corrections. Care is taken to avoid double counting, as the charge radius represents vertex corrections not included in the running of $\sin^2 \theta_W$.

## 3. RESULTS AND DISCUSSION

The maximum recoil kinetic energy depends on the energy of the incident neutrino, and the momentum transfer is a direct consequence of the recoil energy (Fig. 1).

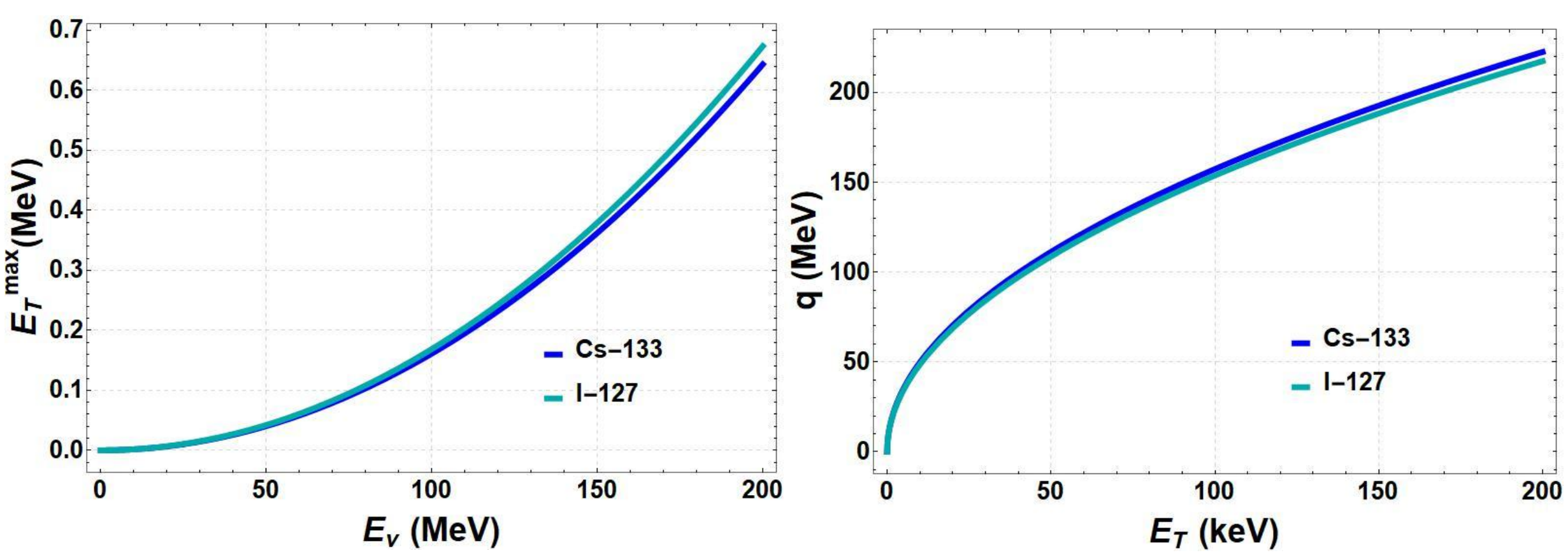

*Fig. 1: a. (Left) Maximum Recoil Kinetic Energy as a Function of Incident Neutrino Energy b. (Right) Momentum Transfer as a function of Recoil Kinetic Energy*

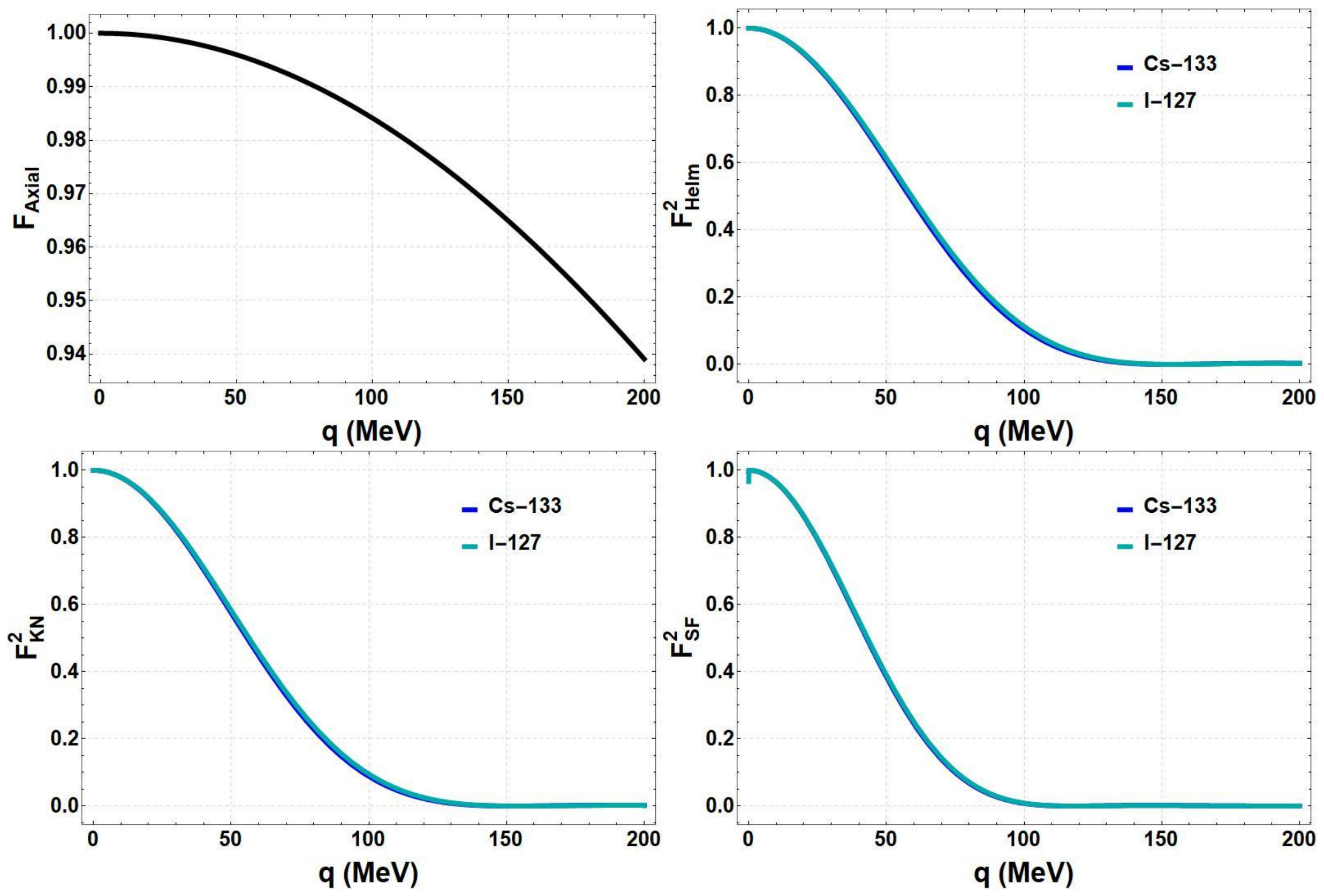

*Fig. 2: a. (Top-left) Nucleon Axial Form Factor b. (Top-right) Nuclear Helm Form Factor c. (Bottom-left) Nuclear KN Form Factor d. (Bottom-right) Nuclear Symmetrized-Fermi Form Factor*

Fig. 2 clearly shows that the nucleon axial form factor and different parameterizations of nuclear form factors decay with increasing momentum transfer. The SF form factor decreases faster than the others. In Fig. 3, neutrino spectral functions are plotted with neutrino energies. These neutrino spectra are used as inputs in our calculations. The plot in Fig. 4 gives us insight into the comparison of neutrino–nucleus cross sections through various channels (CEνNS, incoherent, EM, spin-dependent) for Cs-133 and I-127 nuclei. It shows that as energy increases, the enhancement in the cross section due to incoherent and other effects becomes dominant over CEνNS. For small energies, CEνNS dominates, consistent with literature and experiments. For example, at $E_\nu = 150\, MeV$, CEνNS gives $\sigma_{Cs} \approx\ 3.9 \times 10^{-38}\ cm^2, \sigma_I \approx\ 3.7 \times 10^{-38}\ cm^2$ but by including the effects we get $\sigma_{Cs,I} = 1.07\sigma_{CE\nu NS}$. If we move to higher energies, the coherence is completely lost and the extra effects dominates. For Example, at $E_\nu = 500\, MeV$, cross-section reads $\sigma_{Cs} = 6.07\sigma_{Cs-CE\nu NS}, \sigma_I = 6.2\sigma_{I-CE\nu NS}$. Now it is clear that in CsI [Na] scintillator (used in the first CEνNS detection by the COHERENT experiment at the Spallation Neutron Source), the excitation gammas due to incoherent channel and other effects needs to be detected in order to detect high-energy neutrinos. Fig. 4 provides a general discussion of CEνNS and all associated effects with neutrino-nucleus scattering at high and low energies.

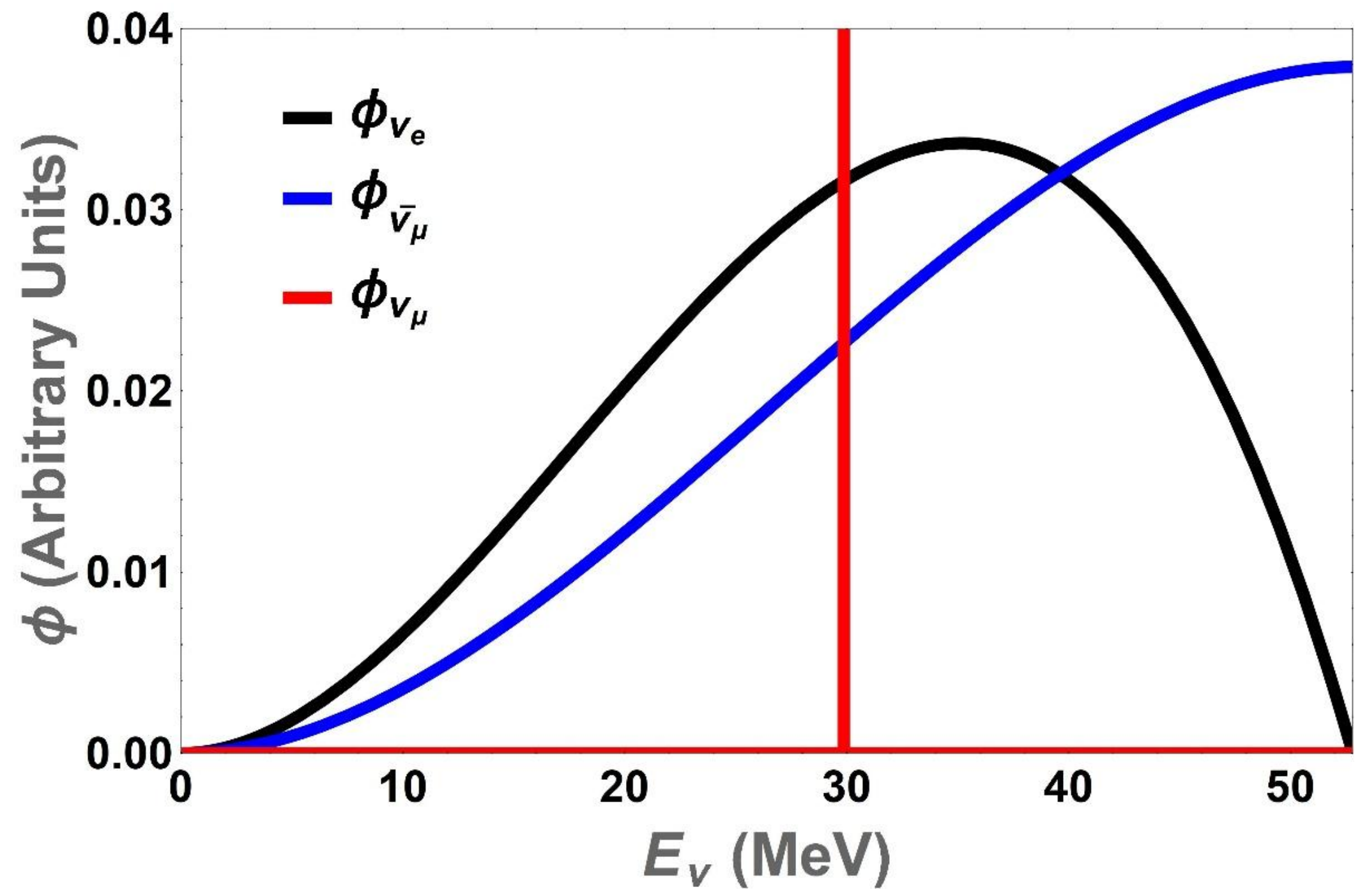

*Fig. 3: Decay-at-Rest (DAR) Neutrino Flux of $\nu_e, \nu_\mu$ and $\bar{\nu}_\mu$*

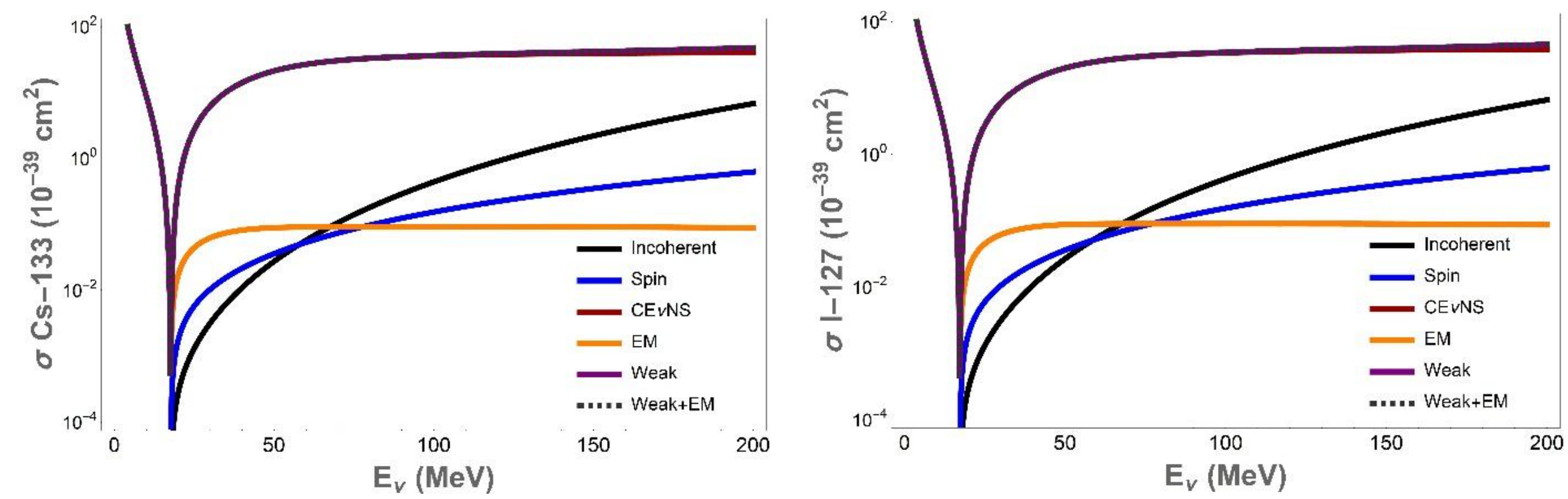

*Fig. 4: Neutrino-Nucleus Cross-sections for Cesium-133 (left) and Iodine-127 (right)*

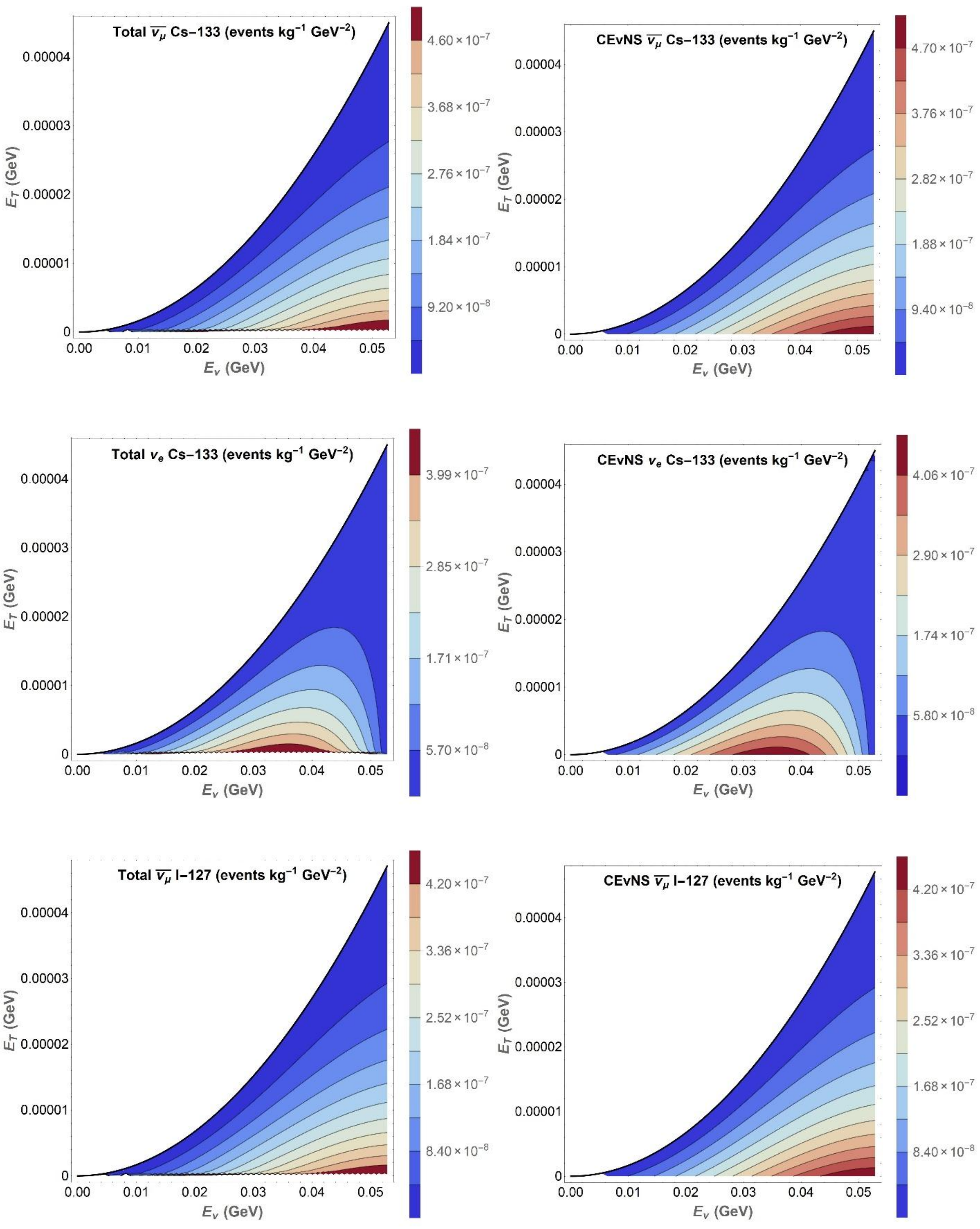

Total $\overline{\nu_\mu}$ Cs–133 (events kg$^{-1}$ GeV$^{-2}$)
CEvNS $\overline{\nu_\mu}$ Cs–133 (events kg$^{-1}$ GeV$^{-2}$)
Total $\nu_e$ Cs–133 (events kg$^{-1}$ GeV$^{-2}$)
CEvNS $\nu_e$ Cs–133 (events kg$^{-1}$ GeV$^{-2}$)
Total $\overline{\nu_\mu}$ I–127 (events kg$^{-1}$ GeV$^{-2}$)
CEvNS $\overline{\nu_\mu}$ I–127 (events kg$^{-1}$ GeV$^{-2}$)
$E_T$ (GeV)
$E_\nu$ (GeV)

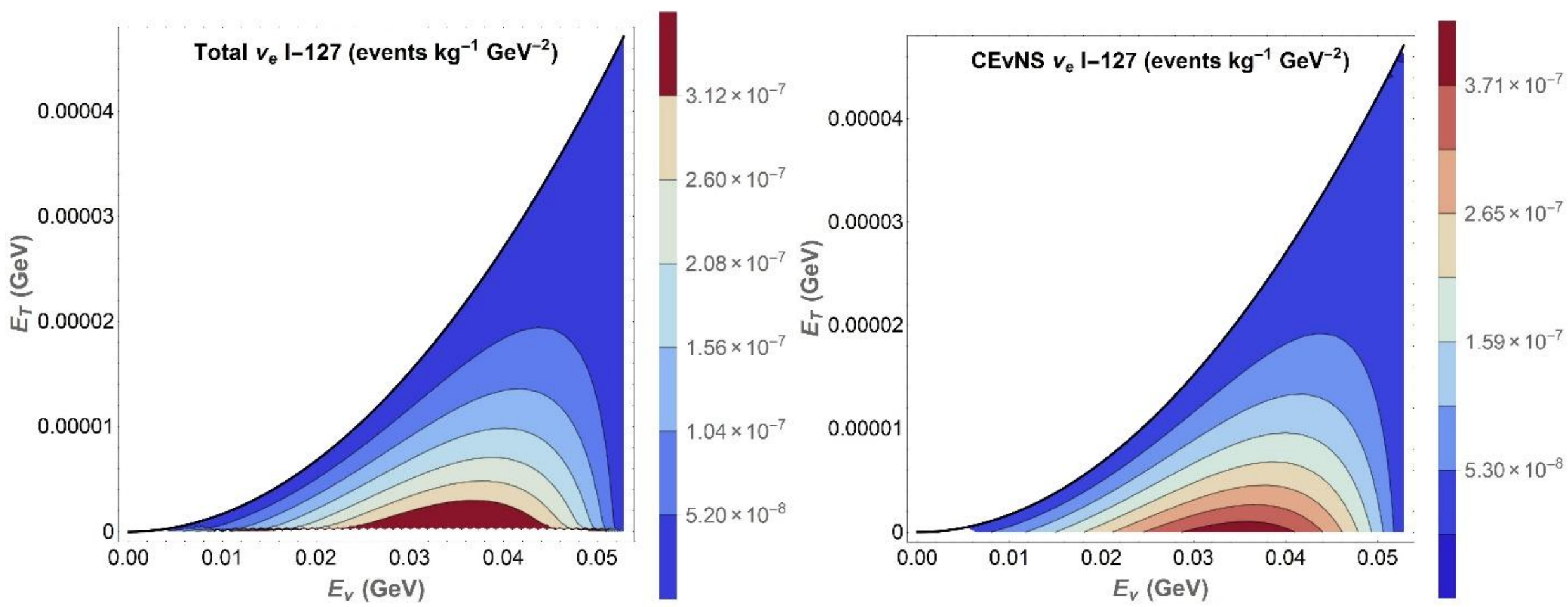

*Fig. 5: Comparison of Expected Double Differential Rate to the CEvNS for Cs-133 and I-127 Nuclei having two flavors of Neutrinos ($\nu_e$ and $\bar{\nu}_\mu$)*

In Fig. 5, the expected double differential rate is greater than that predicted by CEvNS alone due to the inclusion of additional effects at high energies. At low energies, the values remain consistent with CEvNS predictions.

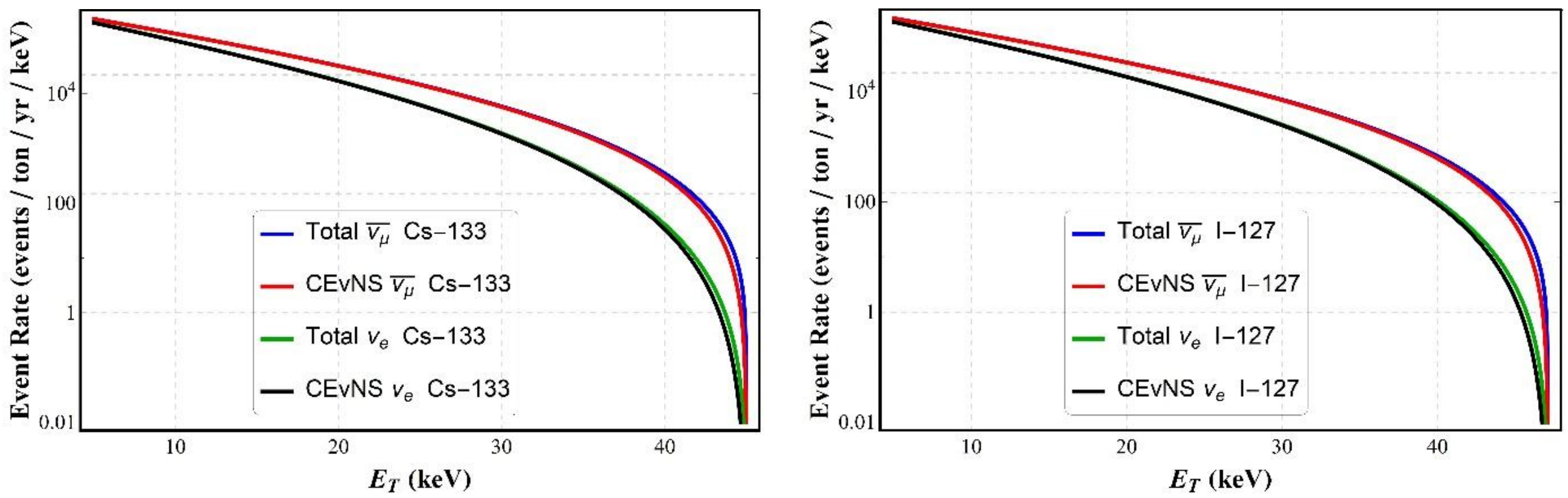

*Fig. 6: Expected Interaction Rate for Cs-133 and I-127 Nuclei having two flavors of Neutrinos ( $\nu_e$ and $\bar{\nu}_\mu$ ) compared with CEvNS*

Fig. 6 shows the interaction rate for $\nu_e$ ,$\bar{\nu}_\mu - N$ scattering for DAR neutrinos. This approach can be generalized to other neutrino sources, but here we used DAR neutrinos to compute interaction rates. We used a realistic recoil-energy threshold for the CsI detector and assumed that the DAR flux fully reaches the detector (idealized case of 100% detection efficiency). The total interaction rate is slightly larger than the CEvNS prediction at higher energies due to the inclusion of additional channels. Even for DAR neutrinos, this effect remains small but non-negligible. Taking a specific example for DAR neutrinos, the expected Interaction rates come out to be:

$$K_{\nu_e-I}\,(40\;keV) = \;1.14IR^{CE\nu NS}_{\nu_e-Cs}, K_{\bar{\nu}_\mu-I}\,(40\;keV) = \;1.10IR^{CE\nu NS}_{\bar{\nu}_\mu-Cs}$$

$$K_{\nu_e-Cs}\,(40\;keV) = \;1.22IR^{CE\nu NS}_{\nu_e-Cs}, K_{\bar{\nu}_\mu-Cs}\,(40\;keV) = \;1.15IR^{CE\nu NS}_{\bar{\nu}_\mu-Cs}\;,$$

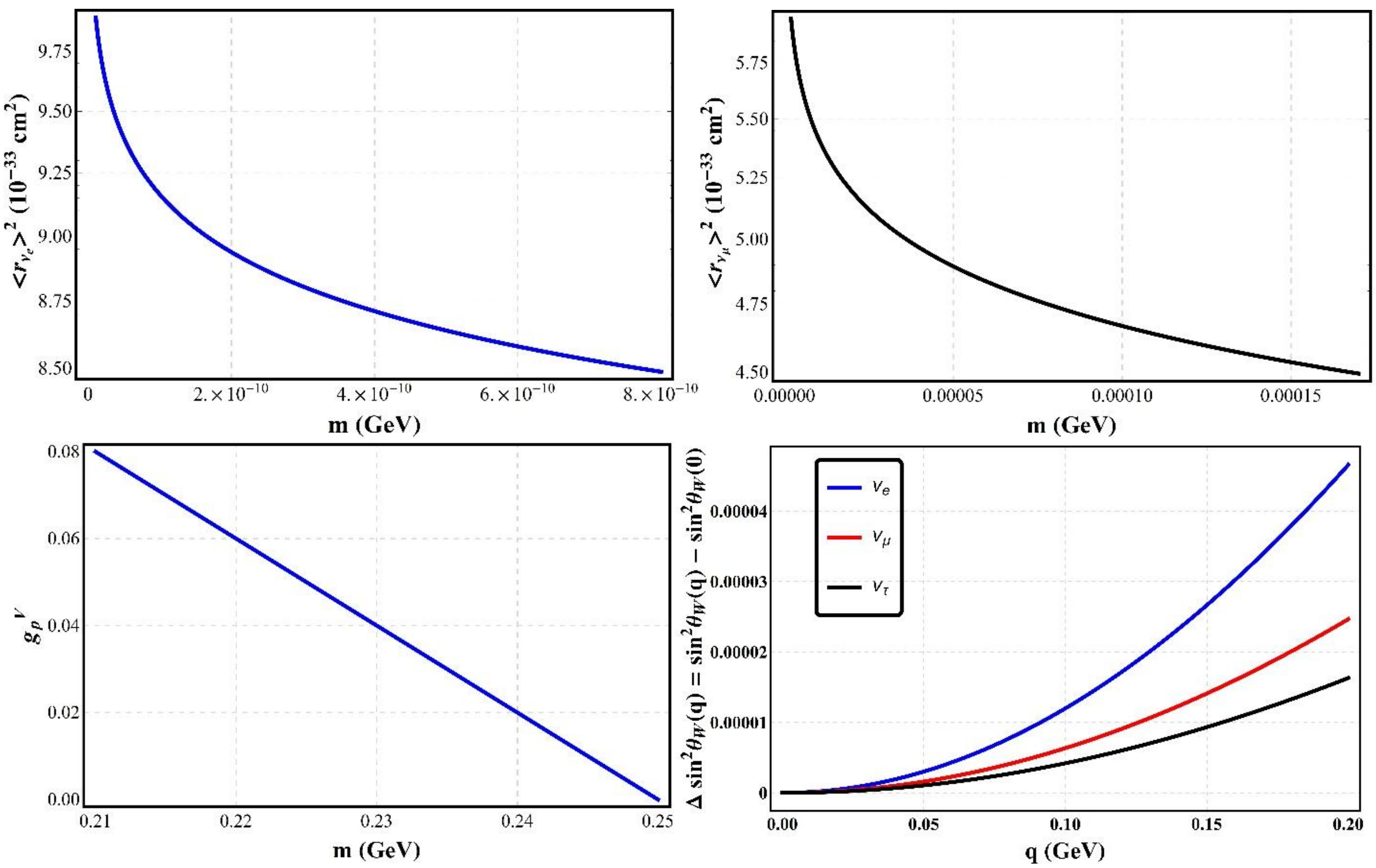

*Fig. 7: a-b. (Top) BSM Neutrino Charge Radii dependence upon masses c. (Bottom-left) Proton Vector Coupling dependence upon $sin^2\,\theta_w$ d. (Bottom-right) Shift in $sin^2\,\theta_w$ by inclusion of momentum dependence and neutrino charge radii*

Fig. 7 shows the neutrino charge radius, the proton vector coupling, and the corresponding shift in the effective weak mixing angle $\sin^2\theta_W$. In many calculations, a reference value $\sin^2\theta_W = 0.2312$ is used; in this work we incorporate its momentum-transfer dependence through our effective parameterization, this modifies the weak charge and, consequently, the neutrino–nucleus cross sections. We include these effects together with the neutrino charge radius contribution (as defined in Sec. 2) and examine their impact on expected cross sections and interaction rates. Neutrino charge radius decreases with the increasing mass of neutrino (BSM). We incorporate these effects instead of using traditional values of $\sin^2\theta_w$ and Proton vector coupling.

Fig. 8 compares neutrino–nucleus cross sections for $^{133}$Cs and $^{127}$I across different interaction channels (CEνNS, incoherent, electromagnetic, and spin-dependent), with electroweak effects incorporated through the momentum-transfer dependence of $sin^2\theta_W$ and neutrino charge radius corrections, which modify the effective vector couplings entering the cross section. At low energies, CEνNS dominates and remains consistent with existing literature and experimental expectations. As an illustrative example, at $E_\nu = 150\,MeV$ , $\sigma^{CEvNS}_{\nu_e-Cs} \approx 4.4\times10^{-38}\,cm^2, \sigma^{CEvNS}_{\bar{\nu}_\mu-Cs} \approx 4.16\times10^{-38}\,cm^2,\ \sigma^{CEvNS}_{\nu_e-I} \approx 4.15\times10^{-38}\,cm^2, \sigma^{CEvNS}_{\bar{\nu}_\mu-I} \approx 3.92\times10^{-38}\,cm^2,\ \ \sigma_{Cs,I} = 1.06\sigma_{CEvNS}$. For higher energies like $E_\nu = 500\,MeV$, cross-section reads $\sigma^{total}_{\nu_e-Cs} = 5.64\sigma^{CEvNS}_{\nu_e-Cs},\ \sigma^{total}_{\bar{\nu}_\mu-Cs} = 5.75\sigma^{CEvNS}_{\bar{\nu}_\mu-Cs},\ \sigma^{total}_{\nu_e-I} = 5.76\sigma^{CEvNS}_{\nu_e-I},\ \sigma^{total}_{\bar{\nu}_\mu-I} = 5.87\sigma^{CEvNS}_{\bar{\nu}_\mu-I}$ .

To validate our results, we compare the order of magnitude of the CEvNS cross sections with established theoretical and experimental studies. The standard description of coherent elastic neutrino–nucleus scattering, first introduced by Daniel Z. Freedman [1], predicts cross sections of order $10^{-39} - 10^{-38}\, cm^2$ in the energy range $E_\nu \sim 50 - 150\, MeV$. This behavior has been confirmed by detailed nuclear calculations [25, 42, 43, 44] and experimentally observed by the COHERENT Collaboration [4, 5]. Our results fall within this expected range, confirming that the CEvNS contribution used as a baseline in this work is consistent with established results. Deviations at higher energies arise from the inclusion of additional interaction channels and electroweak corrections.

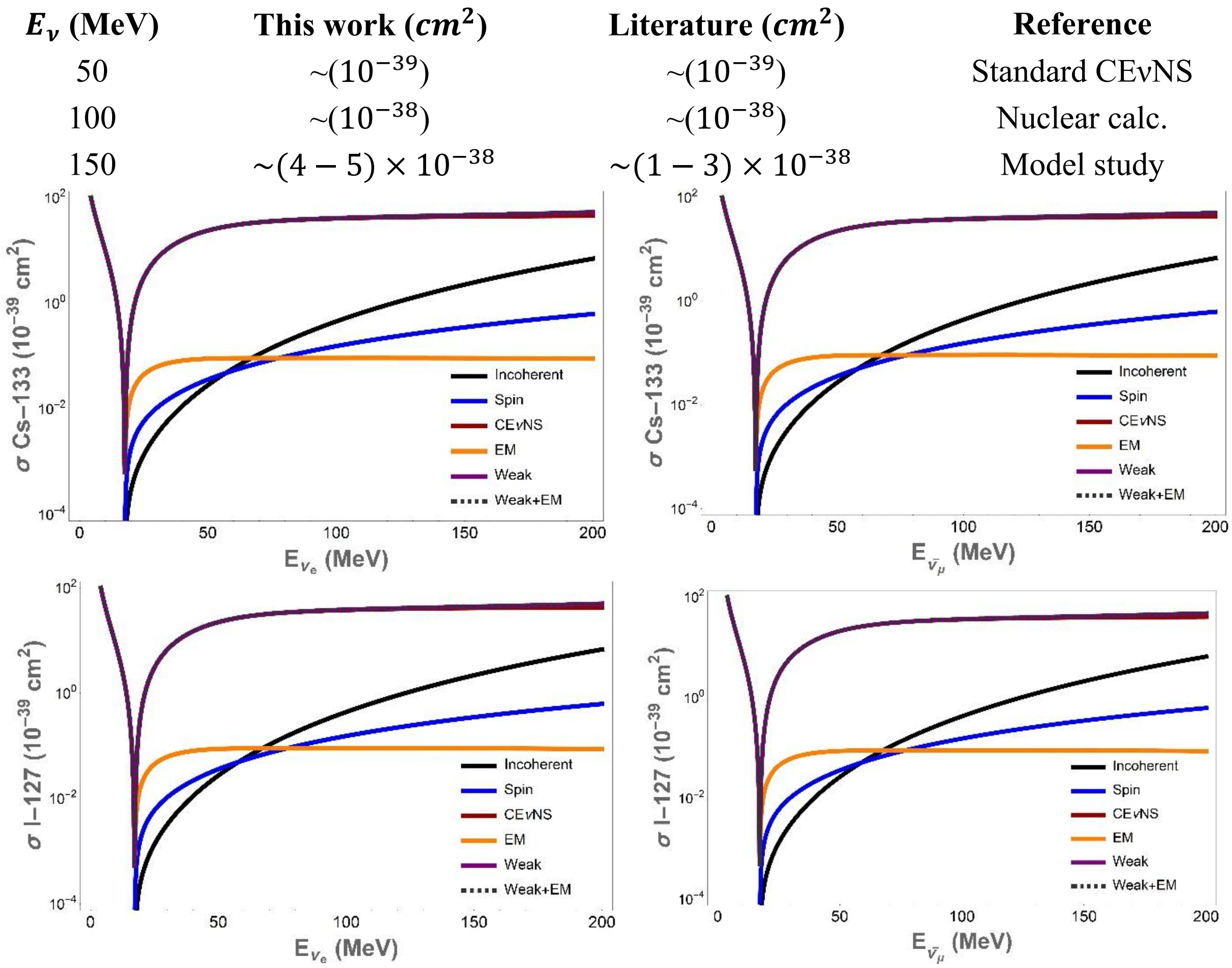

| $E_\nu$ (MeV) | This work ($cm^2$) | Literature ($cm^2$) | Reference |
|---|---|---|---|
| 50 | ~$(10^{-39})$ | ~$(10^{-39})$ | Standard CEvNS |
| 100 | ~$(10^{-38})$ | ~$(10^{-38})$ | Nuclear calc. |
| 150 | $\sim(4-5)\times 10^{-38}$ | $\sim(1-3)\times 10^{-38}$ | Model study |

*Fig. 8: Effective Neutrino-Nucleus Cross-sections for Cesium-133 and Iodine-127*

Effective $\overline{\nu_\mu}$ Cs−133 (events kg$^{-1}$ GeV$^{-2}$)

Effective $\nu_e$ Cs−133 (events kg$^{-1}$ GeV$^{-2}$)

Effective $\overline{\nu_\mu}$ I−127 (events kg$^{-1}$ GeV$^{-2}$)

Effective $\nu_e$ I−127 (events kg$^{-1}$ GeV$^{-2}$)

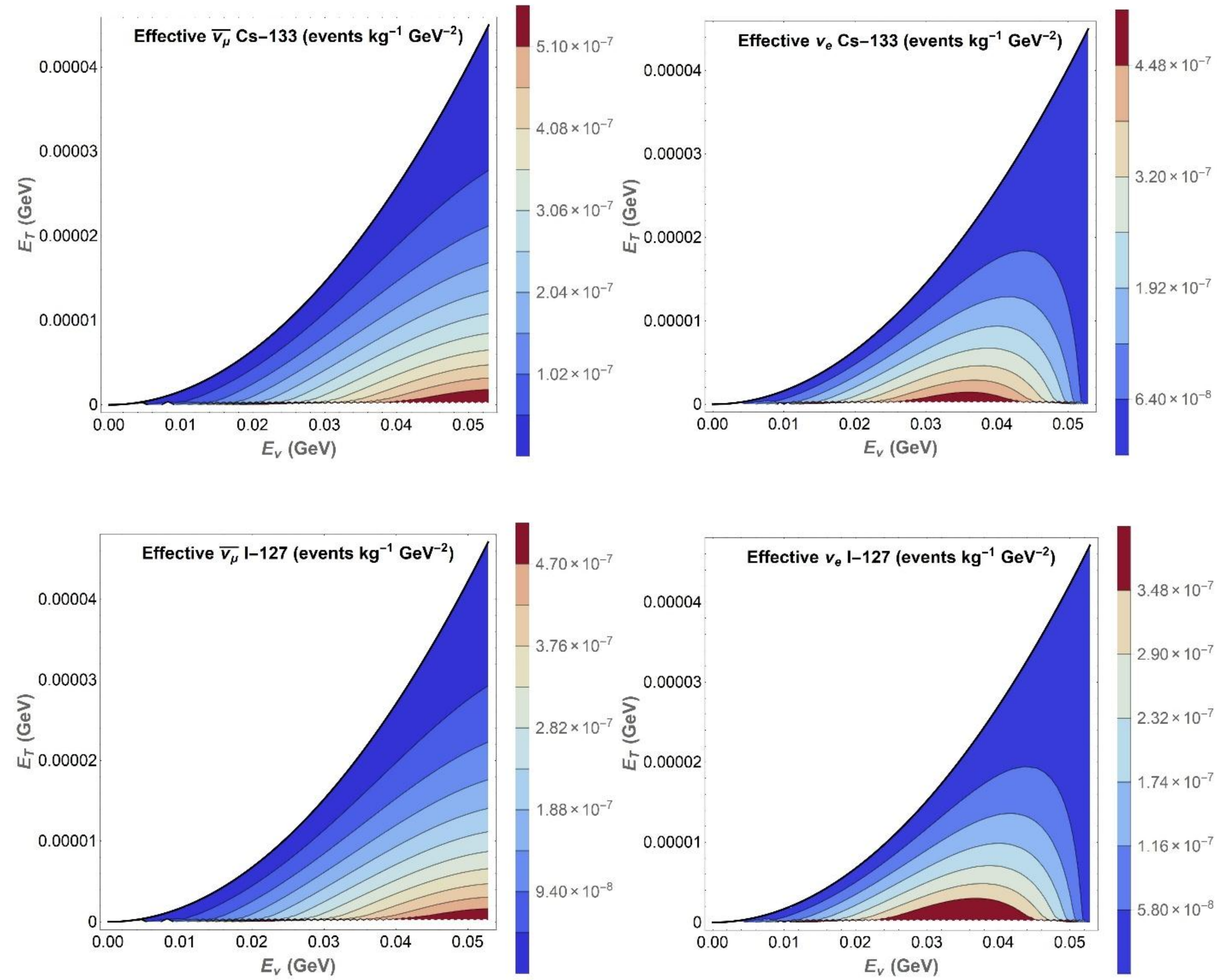

*Fig. 9: Comparison of Expected Double Differential Rate to the CEvNS for Cs-133 and I-127 Nuclei by the inclusion of $\Delta \sin^2 \theta_w$*

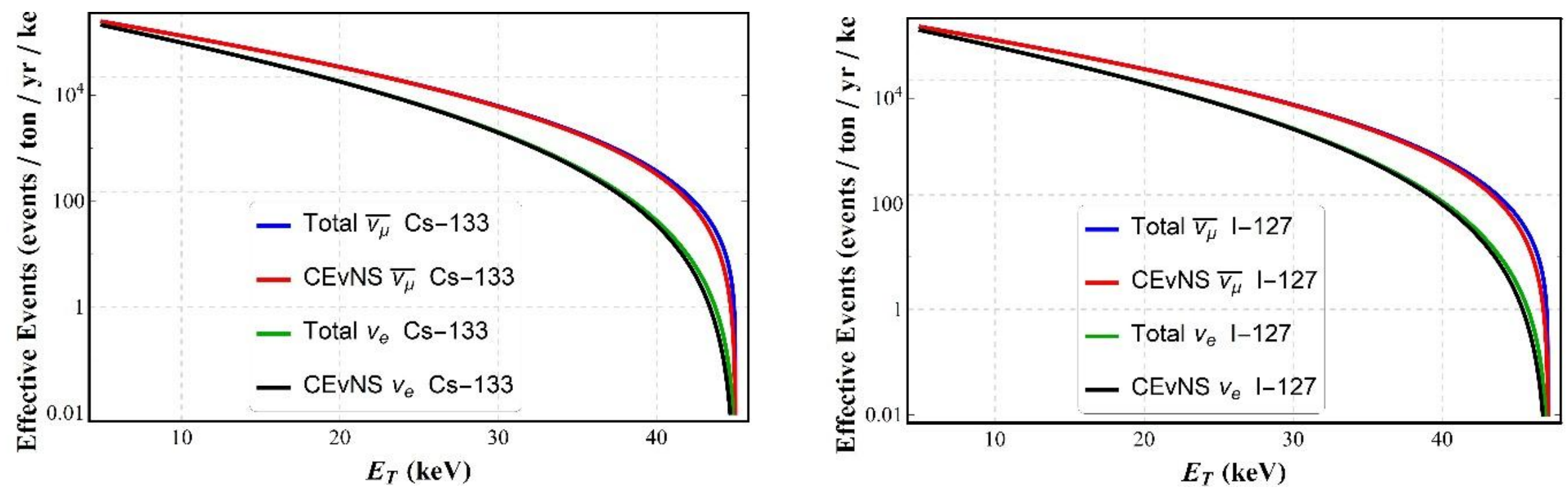

*Fig. 10: Expected Interaction Rate for Cs-133 and I-127 Nuclei compared with CEvNS by the inclusion of $\Delta \sin^2 \theta_W$*

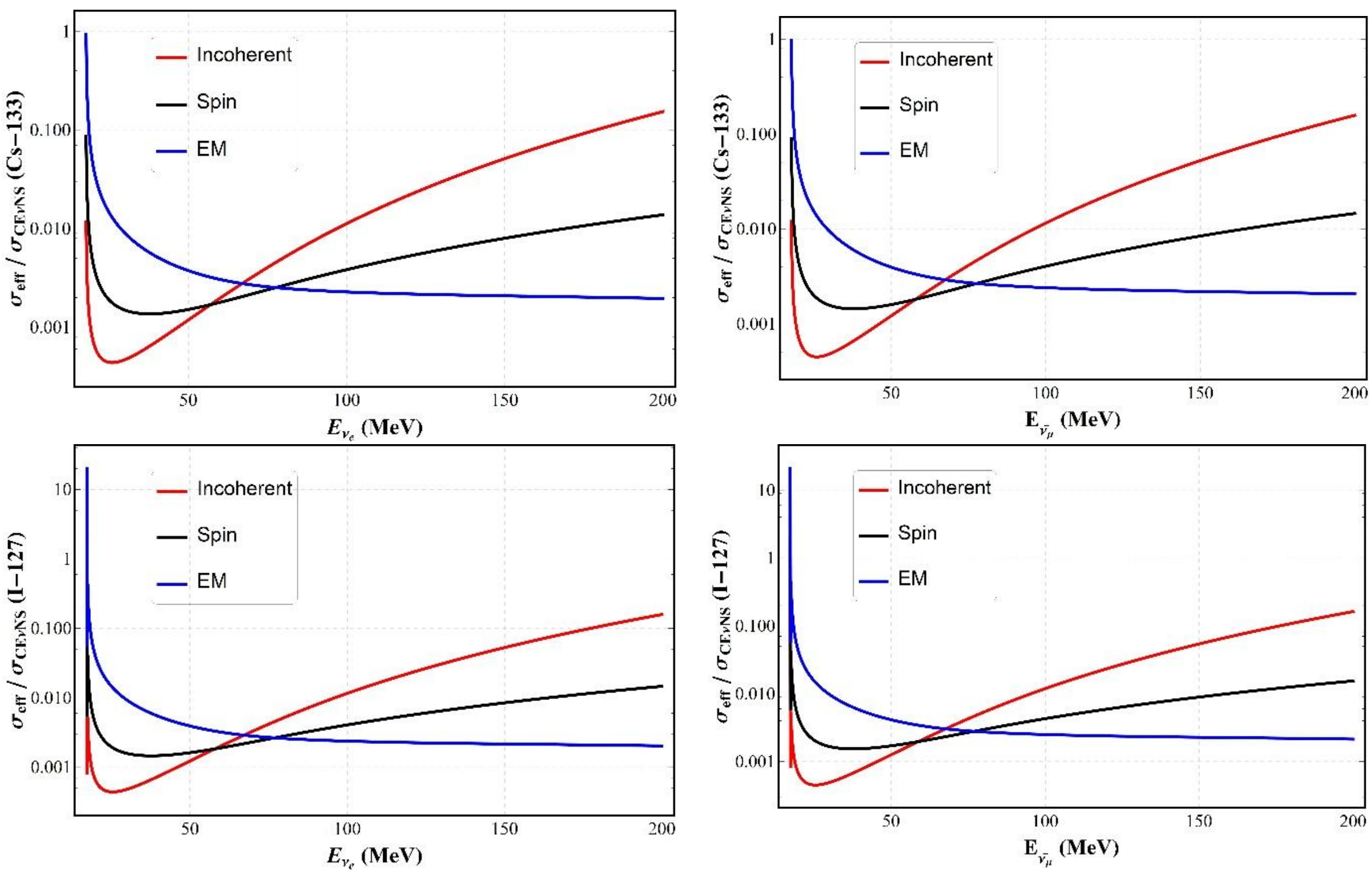

*Fig. 11: Ratios of Cross-sections to CEvNS*

Fig. 9 shows that the expected double differential rate including neutrino charge radius effects is larger than that predicted by CEvNS alone at higher energies. At low energies, both models coincide, confirming CEvNS dominance.

Fig. 10 shows that inclusion of the neutrino charge radius leads to a modest enhancement in interaction rates for Cs-133 and I-127, particularly at higher recoil energies. Taking a specific example for DAR neutrinos, this study shows that expected Interaction rates with electroweak effects are:

$$K_{\nu_e-Cs}\,(40\ keV) = \ 1.21IR^{CEvNS}_{\nu_e-Cs}, K_{\bar{\nu}_\mu-Cs}\,(40\ keV) = \ 1.14IR^{CEvNS}_{\bar{\nu}_\mu-Cs}\ ,$$
$$K_{\nu_e-I}\,(40\ keV) = \ 1.14IR^{CEvNS}_{\nu_e-Cs}, K_{\bar{\nu}_\mu-I}\,(40\ keV) = \ 1.10IR^{CEvNS}_{\bar{\nu}_\mu-Cs}$$

The neutrino charge radius inclusion increases the accuracy of the results at both low and high energies. At low energies, CEvNS dominates. At higher energies, the ratio of total to CEvNS cross section decreases as additional channels become relevant. This energy dependence is important for modeling neutrino–nucleus scattering across different regimes.

Figure 11 presents the ratio of the effective cross sections to the CEvNS cross section. The ratio decreases gradually with increasing neutrino energy, indicating that CEvNS remains the dominant contribution at low energies, while additional interaction channels become increasingly relevant at higher energies. It is evident from the figure that the neutrino charge radius does not significantly affect either $\sigma_{\rm EM}$or $\sigma_{\rm Spin}$. Consequently, the dependence on the neutrino charge radius is relevant only within a specific energy range, where it provides a more accurate description of the higher-energy regime.

Although individual aspects of neutrino–nucleus scattering—such as coherent and incoherent contributions, axial effects, and electroweak corrections—have been extensively investigated in the literature, the present work provides a unified and self-consistent framework in which all these effects are incorporated simultaneously. In particular, we combine coherent, incoherent, electromagnetic, and spin-dependent channels within a consistent implementation of electroweak effects, including the momentum-transfer dependence of the weak mixing angle and the neutrino charge radius.

This approach enables a systematic investigation of neutrino–nucleus interactions across a broad energy range, highlighting the transition from the coherent to the incoherent regime and quantifying the impact of sub-leading effects on observable cross sections and interaction rates in realistic detector materials such as CsI. To the best of our knowledge, such a combined and consistently implemented analysis in this energy regime has not previously been presented in this form.

The present study is primarily theoretical and phenomenological in nature, and a complete uncertainty propagation has not been performed. The principal sources of uncertainty include nuclear form factors (e.g., neutron distribution parameters), the axial coupling $g_A$(including possible quenching effects), neutrino flux normalization and spectral shape, and detector-related effects such as recoil threshold and detection efficiency. In addition, the phenomenological treatment of electroweak running and the neutrino charge radius introduces a degree of model dependence. Depending on the energy range and nuclear target, these effects may lead to variations ranging from a few percent to several tens of percent. A comprehensive uncertainty analysis and direct fits to experimental data are beyond the scope of the present work and are deferred to future studies.

In this work, neutrino energies up to $E_\nu \sim 200$ MeVare considered. Within this regime, coherent elastic neutrino–nucleus scattering (CEνNS) is known to dominate at low energies, whereas coherence gradually diminishes as the momentum transfer increases. At higher energies, additional processes such as incoherent scattering, nuclear excitation, and possible nucleon emission may become important.

The present analysis partially accounts for these effects through the inclusion of incoherent, spin-dependent, and electromagnetic contributions. However, a complete treatment of all inelastic channels and nuclear breakup processes would require a more sophisticated nuclear-structure model. Therefore, the results obtained at higher energies should be interpreted as an effective description that captures the dominant physical trends rather than as a fully comprehensive treatment of all possible reaction channels.

## 4. CONCLUSION

The calculations presented in this work demonstrate that a consistent treatment of neutral-current neutrino–nucleus scattering off $^{127}$I and $^{133}$Cs requires going beyond the coherent elastic approximation, even at relatively modest momentum transfers. By incorporating incoherent excitations, spin-dependent axial contributions arising from the non-zero ground-state nuclear spins, and electroweak corrections such as the momentum dependence of $sin^2\theta_W$ and the neutrino charge radius, the total cross section acquires non-negligible and systematically quantifiable modifications.

As the neutrino energy increases, incoherent processes become progressively more significant and may even dominate the total response for lighter nuclei. Folding these cross sections with decay-at-rest neutrino spectra yields interaction rates of the order of 0.1 events per kilogram per year near a 40 keV recoil threshold, exceeding predictions based solely on coherent scattering.

These results provide a more complete description of the scattering process in CsI[Na] detectors and establish a framework for interpreting deviations from simplified CEvNS expectations. Although the predicted spectra exhibit qualitative agreement with observations from the COHERENT experiment, a detailed comparison with experimental data—including detector effects and systematic uncertainties—lies beyond the scope of the present work.

The role of nuclear spin is particularly important for the odd-A nuclei considered here ($^{127}$I with $J^\pi = 5/2^+$and $^{133}$Cs with $J^\pi = 7/2^+$). In these systems, axial contributions—absent in even–even nuclei—become increasingly relevant with growing momentum transfer, modifying both the magnitude and spectral shape of the cross section. This underscores the importance of incorporating spin-dependent effects in realistic modeling of heavy nuclear targets.

Beyond terrestrial experiments, these results may also have implications for neutrino transport in dense astrophysical environments such as supernova cores, where the interplay between coherent and incoherent scattering channels can influence energy deposition and neutrino propagation.

The present framework, based on analytical expressions and empirical form-factor parameterizations, provides a practical and internally consistent approach for estimating neutrino–nucleus cross sections in the low-to-intermediate energy regime. Nevertheless, a more comprehensive treatment of nuclear structure effects—particularly for axial responses and inelastic channels—would further enhance the robustness of the predictions. Such refinements, together with systematic comparisons to experimental data, constitute important directions for future investigation.